\documentstyle[amsfonts,amssymb,epsfig,12pt]{article}

\newcommand{\sech}{{\mathrm{sech}}}
\begin{document}

\title{Negative Specific Heat of a \\
	Magnetically Self-Confined Plasma Torus}

\author{Michael K.-H. Kiessling$^*$ and Thomas Neukirch$^\dagger$\\ \\
$^*$Department of Mathematics\\
	 Rutgers, The State University of New Jersey\\
	110 Frelinghuysen Rd., Piscataway,  NJ 08854, USA \\
	miki@math.rutgers.edu\\
$^\dagger$School of Mathematics and Statistics,\\
	University of St. Andrews, St. Andrews KY16 9SS, UK\\
	thomas@mcs.st-and.ac.uk}

\maketitle

%%%%%%%%%%%%%%%%%%%%%%%%%%%%%%%%%%%%%%%%%%%%%%%%%%%%%%%%%%%

\begin{abstract}
\noindent
	It is shown that the thermodynamic maximum entropy principle 
predicts negative specific heat for a stationary magnetically self-confined 
current-carrying plasma torus.
	Implications for the magnetic self-confinement of fusion plasma 
are considered. 
\end{abstract}

\bigskip\bigskip

{\centerline{{\textbf{CLASSIFICATION:}}\\
			PHYSICAL SCIENCES\\
			 Applied Mathematics}}

%%%%%%%%%%%%%%%%%%%%%%%%%%%%%%%%%%%%%%%%%%%%%%%%%%%%%%%%%%%

\vfill
\hrule
\smallskip
\noindent
\copyright{2002} The authors. Reproduction of this  article, in its
entirety, for non-commercial purposes is permitted.

\newpage
                  
%%%%%%%%%%%%%%%%%%%%%%%%%%%%%%%%%%%%%%%%%%%%%%%%%%%%%%%%%%%%
%
% body of paper here
% 

	The goal of the controlled thermonuclear fusion program is 
to make the energy source that powers our Sun available to human society. 
	Deep in the Sun's interior, favorable conditions for the 
quasi-stationary 
nuclear burning of the solar plasma prevail as a result of the immense 
gravitational self-forces that keep this huge accumulation of matter together. 
	Since gravitational self-confinement is not operative at the 
reactor and laboratory scale, alternate means of confinement have to 
be employed to achieve sufficiently high plasma densities and temperatures 
in a reactor. 
	In the perhaps most prominent stationary fusion reactor scheme,
the tokamak, strong electric ring currents are induced in an electrically 
neutral plasma to achieve axisymmetric magnetic self-confinement 
in a rotationally invariant toroidal vessel ${\cal T}$.
	In a torus with sufficiently large major axis, such a magnetic 
self-confinement mimics gravitational self-confinement on account of the 
Biot-Savart law, according to which in a system of parallel current 
filaments all filaments attract each other magnetically with the same 
force law as would be the case gravitationally in a system of parallel 
mass filaments. 
	What makes the magnetic forces more attractive (in the double
sense of this phrase) than gravity for laboratory purposes is their 
very much bigger coupling constant ($\approx 10^{40}v^2/c^2$ for two 
electrons moving on parallel trajectories with speed $v$ as 
measured in the laboratory; note that the even stronger ($v^2/c^2$
is replaced by 1)  electrostatic repulsive forces between the same 
two electrons are very effectively screened in a neutral plasma and 
are traditionally neglected to a good approximation).
	Of course, the analogy does not extend to all aspects of
plasma self-confinement.
	In particular, the solenoidal-vectorial character of stationary 
current densities  necessitates the toroidal topology of magnetic
self-confinement, whereas gravity not only allows but manifestly 
prefers spherical confinement over toroidal.
	Unfortunately, axisymmetric toroidal magnetic self-confinement is
not known for its stability either. 
	Although major efforts are devoted to the stabilization of the 
plasma configuration, a vast reservoir of instabilities capable of destroying 
the confinement has dramatically slowed down the development of a operating 
tokamak fusion reactor.
	
	Matters are not exactly helped by the fact that our theoretical
understanding of the physics on the various  space-time scales that govern 
magnetic plasma confinement is still quite incomplete.
	In particular, while the solenoidal character of the magnetic 
induction together with the axisymmetry and stationarity of the law of 
momentum balance tell us that the poloidal magnetic flux function $\Psi$ 
and the toroidal current density $j$ must satisfy some local functional 
relation ${\cal R}(\Psi,j,r)=0$ (in which $r$ is the cylindrical distance 
from the axis of symmetry) which  turns Ampere's law  into some in general
nonlinear elliptic partial differential equation for $\Psi$ (known in the 
fusion literature as the Grad--Shafranov equation), the actual relation
${\cal R}$ is not fixed (except for the explicit $r$-dependence 
in ${\cal R}$).
	Some information about ${\cal R}$ should be contained 
in the law of energy balance between current drive (through an applied 
toroidal $\emph{emf}$ and other means), ohmic heating and, ultimately, 
radiation losses.
	Unfortunately only the so-called classical and neo-classical 
transport coefficients have been computed in some detail \cite{balescuBOOKS}
whereas the small scale turbulent dissipation mechanisms in a tokamak plasma 
remain a largely challenging open problem.
	In this situation, theoreticians have been forced to rely on fair
judgment and good taste when guessing some additional principle(s) that would
effectively complete the characterization of the stationary magnetically 
self-confined plasma torus in a tokamak.

	A sizeable fraction of the literature employs a linear 
approximation of ${\cal R}$ to get $j\propto\Psi$, rendering the 
Grad--Shafranov equation linear, and some improved-accuracy modeling
uses a third-order polynomial approximation of ${\cal R}$ \cite{braams}. 
	Subsequently  the filter of linearized dynamical stability analysis, 
based mostly on macroscopic magnetofluid theory and mesoscopic kinetic theory,
is applied to sort out unstable configurations. 
	While this approach has met with a certain limited success,
one does not learn what the approximations are approximate to.
	Over the years a number of plasma theorists \cite{exppsilita} have 
argued that an equilibrium thermodynamics-inspired maximum entropy principle 
with a few global dynamical constraints\footnote{Since \emph{any} 
	probability density maximizes the entropy 
	relative to itself, a stationary plasma torus is necessarily
	a maximum entropy configuration under \emph{some} constraints. 
	What makes the maximum entropy proposal  non-empty is the insistence
	on only a few global natural dynamical constraints.}
should give answers close to the truth.
	In essence the various formulations in \cite{exppsilita} give
for ${\cal R}$ the answer $j\propto\exp(\Psi)$, which leads to a nonlinear 
Grad--Shafranov equation that may have more than one solution $\Psi$, 
depending on the domain ${\cal T}$, the boundary conditions for 
$\Psi$, and the values of the physical parameters of the problem.
	In addition, this approach provides a global stability criterion
within the class of axisymmetric states satisfying the same dynamical 
constraints.
	Only those solutions that maximize the relevant 
relative entropy functional will be globally stable.

	Since in non-equilibrium statistical mechanics the maximum entropy 
principle has not acquired a status anywhere near as fundamental as in 
equilibrium statistical mechanics, it is mandatory to register some arguments 
in its favor for the case at hand.
	Thus, the relation $j\propto \exp(\Psi)$ has been shown to be 
almost universally singled out also by a truly dissipative Fokker--Planck 
approach to stationary magnetically confined plasma \cite{mkjllA,mikiHABIL}. 
	The perhaps most compelling reason to give it serious considerations,
however, is the successful application of the maximum entropy approach to the 
physically distinct but mathematically quite similar problems of stationary 
planar incompressible flows,\footnote{The first qualitative predictions 
		based on statistical mechanics of the Hamiltonian system of 
		$N$ point vortices were made in \cite{LARS}.
		The quantitative evaluation began with \cite{montgomeryA}; 
		impressive agreement with simulated flows is reported in 
		\cite{montgomeryB}. 
		Its mathematical rigorous foundations are by now almost
		complete, the latest word being \cite{mkjllB};
		see \cite{miki2000} for a review.
		More recently a formulation based directly on 
		continuum vorticity has gained much ground; 
		see \cite{EHTb} for a state-of-the-arts report.}
where the vorticity plays a r{\^{o}}le closely analogous to the
current density in the plasma torus, and the strongly magnetized 
pure electron plasma in a circular cylinder,\footnote{In the guiding 
	center approximation the dynamics of this plasma system is
	identical to that of $N$ point vortices \cite{LIN}. 
	Statistical mechanics 
	in the corotating frame predicts that at high enough effective
	energies the nonlinear $m=1$ diocotron mode has higher 
	entropy than any other configuration with the same energy 
	and angular momentum \cite{smithoneil}, in accordance with
	remarkable real experiments \cite{malmbergetal}.}
where the charge density plays that r{\^{o}}le.
	In this spirit, we have conducted a thorough investigation 
of the thermodynamic-type maximum entropy approach to the magnetically 
self-confined stationary plasma torus \cite{mikiHABIL}. 

	A most curious finding of the study \cite{mikiHABIL} is that
the gravity-inspired toroidal magnetic plasma self-confinement 
scheme inherits from the stars their gravo-thermal
\emph{negative specific heat}.\footnote{Ref. \cite{LL}, pp.\ 60-63,
		explains why a homogeneous piece of ``everyday matter'' 
		must have positive specific heat. 
		See p.\ 62 of the same reference for why those 
		arguments do not rule out negative specific heat 
		in an isolated gravitating system.
		Indeed, the virial and the equipartition theorems imply 
		that in a spherical equilibrium system the energy is 
		distributed -2:1 between gravitational and kinetic. 
		A decrease in total energy $E$ of a gravitational equilibrium 
		gas ball will increase its thermal energy. Such a system 
		grows hotter while losing energy through, say radiation. 
		Negative specific heat in self-gravitational perfect
		gases is evaluated quantitatively already in \cite{EMDEN} 
		and is further discussed in \cite{redgiants}; however, 
		none of these configurations with negative specific heat 
		is thermodynamically stable though some are metastable. 
		Thermodynamically stable self-gravitating configurations 
		with negative specific heat can occur when the Newtonian 
		$-1/r$ singularity is stabilized either as in quantum 
		mechanics \cite{Thirring} or in classical hard balls 
		systems \cite{SKS}.}
	This result is a little surprising, for it follows from what we 
said earlier that the plasma torus should actually more closely mimic
a cylindrical caricature of a star, and the specific heat of a 
`cylindrical maximum entropy star' \cite{ALYb} and its plasma physical 
clone, the cylindrical Bennett pinch \cite{bennett}, is non-negative! 
	The existence of a maximum entropy plasma torus with negative 
specific heat is therefore a truly nontrivial fact.
	The purpose of this note is to point out some potentially important 
consequences of our finding for plasma physical 
applications.\footnote{Recently, the existence and importance of negative 
		specific heat was also reported for the diocotron mode 
		of the guiding center plasma 
		\cite{smithoneil} alias point vortex gas, and for certain 
		vorticity structures in geostrophic flows \cite{EHTb,TMHDB}.
		However, very different from the gravo-thermal type 
		negative specific heat that we report here to be a 
		characteristic also of the magnetically self-confined 
		plasma torus, the negative specific heat of these 
		quasi-particle systems does \emph{not} couple to the 
		thermal motion of the underlying physical particle systems, 
		which is evident from the fact that these quasi-particle 
		systems also exhibit negative temperature \cite{LARS}.}
	To pave the way for the discussion we first describe the model and
our results.
\smallskip

\noindent{\textbf{The model}}

\noindent
	Since the whole problem is rotationally invariant,
we work with conventional cylindrical coordinates $r,\theta,z$. 
	The magnetic induction field decomposes accordingly as 
${\mathbf{B}} = {\mathbf{B}}_T + {\mathbf{B}}_P$,
where ${\mathbf{B}}_T \| {\mathbf{e}}_\theta$ 
and ${\mathbf{B}}_P \perp {\mathbf{e}}_\theta$.
	In an actual tokamak, the toroidal component ${\mathbf{B}}_T$
and a part, ${\mathbf{B}}_0$, of the poloidal component are externally 
generated harmonic fields that serve the purpose of azimuthal and  
radial stabilization.
	The total poloidal induction 
${\mathbf{B}}_{P} = \nabla\Psi \times \nabla\theta$
is the sum of ${\mathbf{B}}_0$ and a component which is 
generated by the electric plasma current density vector,
$j{\mathbf{e}}_\theta$, via the toroidal Amp\'ere's law
\begin{equation}
-r\nabla\cdot\left(r^{-2}\nabla \Psi\right)  = 4{\pi}c^{-1} j.
\label{amperelaw}
\end{equation}
	In a similar manner one decomposes the electric field, 
${\mathbf{E}} = {\mathbf{E}}_T + {\mathbf{E}}_P$, 
where ${\mathbf{E}}_T$ is driving the plasma current while
the poloidal part is determined by Coulomb's law
$\nabla \cdot {\mathbf{E}} = 4\pi \rho$, where $\rho$ is the 
electric charge density. 
	Usually the so-called quasi-neutrality approximation is
invoked, which determines the poloidal electric field in leading order 
through a singular perturbative approach to Coulomb's law.
	To keep matters as simple as possible,  we consider 
an `electron-positron' plasma, which is totally charge symmetric 
with regard to the particle species so that $\rho$ vanishes exactly.
	In that case the poloidal electric field vanishes identically, too,
while the toroidal one is implicitly contained in the electric plasma
current $I= Nq\omega /2\pi$. 
	Here $N$ is total number of plasma particles, 
$q$ the elementary charge, and $\omega$ the mean absolute angular
frequency of a species.
	This settles the electromagnetic part of the model, and we 
turn to the statistical mechanics part. 

	While most works in \cite{exppsilita} are formulated at the 
macroscopic magnetofluid level, they can in essence be recovered from 
the statistical mechanics approach of Kiessling et al. in 
\cite{exppsilita}, which begins with the 
Hamiltonian $N$ particle formulation and takes the kinetic limit. 
	At this kinetic level, the plasma particles are described by
distribution functions on single-particle phase space. 
	We seek those distribution functions which maximize the familiar 
Boltzmann entropy functional under the constraints that the two separating 
integrals of motion, particle number and energy, take prescribed
values $N$ and $E$, and given that the plasma carries a prescribed electric 
current $I$. 
	Since the current is not a separating integral, one has to 
resort to a ruse and prescribe each species' canonical angular
momentum, which in the axisymmetric kinetic limit is a 
separating integral, and subsequently pass from this microcanonical 
angular momenta ensemble to its canonical convex dual, characterized
by prescribed $\omega$, viz. $I$.
	The solutions of the corresponding Euler--Lagrange equations for 
this variational principle are also stationary solutions to the
axisymmetric kinetic equations of Vlasov. 
	Over velocity space the resulting distribution functions are
simply rigidly rotating Maxwellians with temperature 
$T = (k_{\mathrm{B}} \beta)^{-1} >0$, tied to the energy constraint, and
angular frequencies $\pm \omega\ (\propto I)$, microcanonically 
tied to the angular momentum constraints but canonically prescribed.
	This allows one to explicitly integrate over the velocity
space to retain only the effective macroscopic entropy principle for 
the total number density of plasma particles, $n({\mathbf{x}}) = 
\overline{n}(r,z)$, which in our charge symmetric plasma is just 
twice the value of the common space-dependent Boltzmann factor of 
each species' distribution function.
	In effect this renders the entropy a functional of $n$,
\begin{equation}
S(n) 
= 
- k_{\mathrm{B}}{\int} n({\mathbf{x}}) 
\ln\left(\lambda^3_{\mathrm{dB}}n({\mathbf{x}})/2\right)d{\cal T}
+ {\frac{5}{2}}Nk_{\mathrm{B}} \, ,
\label{entropy}
\end{equation}
where $\lambda_{\mathrm{dB}} = h/\sqrt{2{\pi}mk_{\mathrm{B}}T}$ is the 
thermal de Broglie wave length.
	This entropy functional has to be maximized under the constraints 
that $n\geq 0$ is axisymmetric, $\int\! n\, d{\cal T} = N$, and that the 
effective energy functional\footnote{The negative sign in
		front of the magnetic energy in $W$ is due to the canonical
		constraint of prescribed electric current; cf. 
		the negative sign in front of the centrifugal contribution
		to the kinetic energy of a rotating thermal system in
		the co-rotating frame, see Landau-Lifshitz 
		(op.\ cit., pp.\ 71-73).
		Incidentally, those very centrifugal contributions to $W$ 
		are negligible in our plasma and have been omitted.
		Moreover, the toroidal field ${\mathbf{B}}_T$ does not show
		since we consider only axisymmetric configurations
		with toroidal current density.} 
\begin{equation}
W(n) = 
-{\frac{1}{2}\int\!\! \int}
	n({\mathbf{x}})
		K({\mathbf{x}},{\mathbf{x}}^\prime)
	n({\mathbf{x}}^\prime)
d{\cal T}d{\cal T}^\prime  
+ {\frac{3}{2}}N k_{\mathrm{B}}T \, 
\label{energy}
\end{equation}
takes a prescribed value, say $E$.
	Here, $K({\mathbf{x}},{\mathbf{x}}^\prime)=
\left(2{\pi}I/cN\right)^2
G({\mathbf{x}},{\mathbf{x}}^\prime)$,
and $G$ is the Green's function for  
$-\nabla\cdot\left(r^{-2}\nabla  \right)$ in ${\cal T}$
for boundary conditions detailed below.
	With the help of a variant of Moser's corollary \cite{moser} of 
the Trudinger--Moser inequality it can be shown that given $I$, 
the entropy functional $S(n)$ takes its finite maximum on the set of
non-negative axisymmetric densities $n({\mathbf{x}}) =\overline{n}(r,z)$ 
satisfying $\int\! n({\mathbf{x}}) d{\cal T} = N >0$ and $W(n) = E$,
and the maximizer is a regular solution of the Euler--Lagrange 
equation.\footnote{These are quite nontrivial facts. 
		In particular, all this is not true if we
		relax the condition of axisymmetry.}
	We remark that more than one maximizing density function $n$ 
might exist, and in addition the nonlinear Euler--Lagrange equation may 
have other types of solutions.
	We call a solution \emph{$S$ stable} if it is a \emph{global}
maximizer of the entropy (for the given constraints), \emph{ $S$ meta-stable} 
if it is merely a \emph{local} maximizer, and unstable otherwise.
	Of course, only those $S$ stable solutions which exhibit 
magnetic self-confinement are of interest.
\smallskip

\noindent{\textbf{Results}}

\noindent
	Explicitly carrying out the variations and converting the 
Euler--Lagrange equation for $n$ into an equation for $\Psi$, using 
$\Psi({\mathbf{x}}) = c^{-1} \int G({\mathbf{x}},{\mathbf{x}}^\prime)
j({\mathbf{x}}^\prime)/r^{\prime} d{\cal T}^\prime$ and
$j({\mathbf{x}}) = qn({\mathbf{x}}) \omega r$, we
obtain Pfirsch's \cite{pfirsch} nonlinear Grad--Shafranov equation 
\begin{equation}
-\nabla\cdot\left(r^{-2}_{{}}\nabla \Psi\right)  = 8\pi^2 c^{-1}I
\frac{e^{\beta \omega q \Psi/c}} 
{\int\! e^{\beta \omega q \Psi/c}\, d{\cal T}} ,
\label{gradshafranovEQ}
\end{equation}
which is to be solved in the torus ${\cal T}$ 
for the boundary conditions encoded in $G$.
	Solving (\ref{gradshafranovEQ}) is in general only possible
numerically on a computer. 
	However, some explicit analytical control is available if 
one simplifies the actual laboratory geometry somewhat and considers
a torus ${\cal T}$ with rectangular cross section
$
\{r,\theta,z| r_i < r < r_o; \theta\ fixed; 0 < z < H\}.
$
	The poloidal flux function 
$\Psi({\mathbf{x}}) =\overline{\Psi}(r,z)$ is assumed to satisfy 
periodic conditions at the $z$ boundary and to be constant at $r_i$ and 
$r_o$, so that the radial component of ${\mathbf{B}}_{P}$ vanishes
at the inner and outer boundaries of ${\cal T}$. 
	In this setting the harmonic poloidal part is simply a homogeneous
${\mathbf{B}}_0 \| {\mathbf{e}}_z$, which we choose so that 
$\overline{\Psi}(r_i,z) = \overline{\Psi}(r_o,z)$.  
	By gauge freedom we can now even set $\overline{\Psi}(r_{i},z) =0$. 
	Beside the desired self-confined configurations, these
boundary conditions allow also unconfined ones, namely Pfirsch's 
toroidal sheet pinch \cite{pfirsch}, given by the following 
${\mathbf{e}}_z$-invariant solution of (\ref{gradshafranovEQ}),
\begin{equation}
\Psi_{\mathrm{Pf}} (r) = -\frac{2c}{\beta\omega q} \ln 
\frac{\cosh(\kappa^2[2r^2- r_o^2 - r_i^2]/2)}{\cosh(\kappa^2[r_o^2-r_i^2]/2)},
\label{pfirschpinch}
\end{equation}
with $\kappa \in (0,\infty)$ a parameter and
$\beta(\kappa^2)$ given by
\begin{equation}
\beta = 
4c^2q^{-2}N^{-1}\omega^{-2}H \kappa^2\tanh(\kappa^2[r_o^2 - r_i^2]/2)\, .
\label{kappaeq}
\end{equation}
	While these solutions do not describe a magnetically 
self-confined plasma torus, they serve as our jumping off point
for the numerical computations of the confined configurations. 
	Our strategy, which was also contemplated by
K. Schindler,\footnote{Private communication.} 
is to look for ringlike bifurcations from the toroidal sheet
solution (\ref{pfirschpinch}). 
	At an infinite sequence of discrete values
$E_1>E_2>...$, with $E_k\searrow E_\infty = - \pi^2 I^2
(r_o^2-r_i^2)/2Hc^2$, other solutions bifurcate off of the sheet pinch 
sequence, breaking its $z$ invariance.
	The bifurcation points are determined by setting 
$\Psi({\mathbf{x}}) = \Psi_{\mathrm{Pf}}(r) + \epsilon \psi(r,z) 
+ O(\epsilon^2)$,
with $\psi(r_{i,a},z)=0$ and $\psi(r,z+H)= \psi(r,z)$,
and expanding (\ref{gradshafranovEQ}) to first order in $\epsilon$, 
giving the linearized problem
\begin{equation}
-\nabla\cdot\left(r^{-2}\nabla \psi\right) 
+ V\psi - V\langle \psi\rangle = 0 ,
\label{linamperelaw} 
\end{equation} 
where $\langle\psi \rangle = \int\psi (r,z)
V(r)d{\cal T}/\int V(r)d{\cal T}$, and
\begin{equation} 
V(r) = - {8\kappa^4}{\sech^2(\kappa^2[2r^2-r_o^2-r_i^2]/2)}.
\end{equation} 
	By Fredholm's alternative \cite{courant}, the solution 
of (\ref{linamperelaw}) is trivial except for certain discrete
values of $\kappa$ at which the bifurcations occur. 
	We have proved \cite{mikiHABIL} that for our ${\cal T}$ 
all bifurcations off of the sheet pinch are due to modes $\psi_k$, 
$k=1,2,...$, that satisfy $\langle \psi_k \rangle = 0$. 
	The first mode is of the form 
$\psi_1(r,z) = R(r)\cos (2\pi [z-z_0]/H)$,
with $z_0$ arbitrary, and with $R(r)$ satisfying 
\begin{equation}
- {r} \left(
{r}^{-1} R^\prime\right)^\prime
+ (2\pi/H)^2 R + r^2VR =0 
\label{linRode}
\end{equation}
for $R(r_i) = R(r_o)=0$. With realistic domain dimensions
$r_i=1 $, $r_o =\sqrt{2}$, and $H =2$,
a standard Runge-Kutta solver finds the unique
nontrivial solution at $\kappa = \kappa_1= 1.62$, giving
$E_1 = 2.72 \, W_\bullet$, with $W_\bullet = 2\pi^2 r_i I^2/25c^2$. 
	Numerical solutions of (\ref{gradshafranovEQ}), with 
$r_i=1 $, $r_o =\sqrt{2}$, and $H =2$, were then computed with 
a well-tested bifurcation code \cite{zwingmannetal},
based on a continuation method \cite{keller}. 
	Our code reproduced the analytical sheet pinch solution 
and its first bifurcation point in excellent agreement with 
our independently obtained semi-analytical results.
	We then numerically followed the first bifurcating branch that 
emerges from $\psi_1$ to nonlinear amplitudes, where it 
develops into a toroidal ring pinch 
with a double X magnetic structure similar to the 
double X structure in the PDX-PBX Tokamak experiment in
Princeton.\footnote{For design information, see 
	{\em http://www.pppl.gov/oview/pages/pbxm\_design.html}.} 
	Retrospectively, this vindicates our choice of boundary conditions. 

 \centerline{(FIG. 1)}

%\begin{figure}% [t]
%\centering\leavevmode
%\epsfxsize=0.4\textwidth
%\epsfbox{Blines.eps}

%\medskip
%\caption{Poloidal magnetic lines of force of maximum entropy solutions near 
%	the second order phase transition at $E_1 = 2.72 \, W_\bullet$. 
%	Ring pinch (left): $W(n)= 2.34 \, W_\bullet$, 
%				$\beta = 0.29 N/W_\bullet$, $z_0 = 0$;
%	Sheet pinch (right): $W(n)= 3.00 \, W_\bullet$, 
%					$\beta = 0.30 N/W_\bullet$.
%	The toroidal hoop effect is neatly visible.\hfill}
%\label{blines}
%\end{figure}

	Our primary interest is in the energy-entropy diagram. 
	Shown in Fig.\ \ref{svsw} is ${\triangle} S(n)$ versus $W(n)$, where 
${\triangle} S(n) = S(n) - Nk_{\mathrm{B}}\ln\sqrt{
		(4\pi e/N)^5(I/ \hbar c)^6 m^3 r_i^3}$,
with $e$ the Euler number, and with the density function $n$ running along 
the computed bifurcation sequences of ring and sheet pinch.
	At sufficiently high effective energies, Pfirsch's sheet pinch 
is the unique solution of (\ref{gradshafranovEQ}) for the stipulated 
boundary conditions, hence maximizing entropy. 
	Numerically it appears to be 
the case for all $W(n) > E_1=2.72 \, W_\bullet$, see Fig.\ \ref{svsw}. 
	For all $W(n)< E_1$ down to $W(n) = - 0.5\, W_\bullet$ where 
we terminated the computation, the ring pinch  has higher entropy than
the toroidal sheet pinch at the same effective energy. 
	By asymptotic analysis we found that also for $W(n)\searrow -\infty$, 
and by continuity for $W(n)\ll -W_\bullet$, the maximum entropy configuration 
consists of a highly concentrated ring pinch which, in rescaled
coordinates centered at the density maximum, 
converges to Bennett's cylindrical pinch \cite{bennett} as 
$W(n)\searrow -\infty$.
	On the basis of this evidence we surmise that the ring
pinch has maximum entropy for all $W(n)<E_1$, implying its $S$ stability 
in the class of rotationally invariant plasma with effective energy
$W(n)<E_1$ and current $I$. 
	We remark that the first bifurcation off of the toroidal 
sheet pinch into the $S$ stable toroidal ring pinch branch 
is then a symmetry-breaking second-order phase transition.

	It remains to determine the specific heat of the configurations,
which we recall is negatively proportional to the second derivative of 
$S$ with respect to $E$.
	Thus we inspect the curvature of the graphs of the entropy as 
function of energy for the various solutions, given in Fig.\ \ref{svsw}. 
	The graph representing the sheet pinch is concave. 
	However, the graph for the ring pinch is manifestly convex
over the whole computed range of energies $- 0.5\, W_\bullet< W(n) < E_1$.
	We were also able to prove the convexity analytically to second 
order in perturbation theory away from the bifurcation point.
	This confirms what we have announced earlier: 
{\textit{the specific heat of the ring pinch is negative!}}

 \centerline{(FIG. 2)}

%\begin{figure}%[b]
%\centering\leavevmode
%\epsfxsize=0.45\textwidth
%\epsfbox{SvsW.eps}
%\medskip
%\caption{S versus W for toroidal sheet pinch (concave branch)
%	and  ring pinch (convex branch). }
%\label{svsw}
%\end{figure}

\smallskip

\noindent{\textbf{Discussion}}

\noindent
	At last, we discuss the potential implications that our 
finding of gravo-thermal type negative specific heat has for the 
problem of toroidal magnetic self-confinement of plasma. 
	In a gravitationally bound plasma, negative specific heat on one 
hand aids the ignition of nuclear burning in a proto-star by heating it up 
when it loses energy by radiation, but it is also responsible for some more 
spectacular instabilities once the nuclear burning expires, like the onset 
of the red giant structure \cite{redgiants}.
	It would be intriguing enough if the negative specific heat 
of a magnetically self-confined plasma torus should be confirmed
to aid the ignition of nuclear burning in a tokamak.
	For this to be so, one would have to be able to hold
$N$ and $I$ fixed and secure the toroidal invariance (which is what
one wants to achieve anyhow), while $E$ would have to decrease (the
plasma radiation would seem to help in this respect) \emph{slow enough} 
so that one would essentially evolve  along the ring pinch branch 
in Fig.2 to the left, thereby heating up the plasma while pinching
it more strongly.
	This would not seem unwelcome.
	For now, however, energy leakage by radiation is a serious 
problem, while at the same time the \emph{emf} current drive leads to yet 
uncontrolled ohmic heating of the plasma. 
	In this case where $E$ is allowed to fluctuate too widely, the 
negative specific heat will have a very unwanted effect on the confinement. 
	This can be illustrated by considering the temperature rather 
than energy $E$ to be controlled by the competition of ohmic heating and
radiation (still assuming $N$ and $I$ fixed, and toroidal invariance).
	In that case the canonical ensemble determines the stability.
	But microcanonical and canonical ensembles are not equivalent
when the microcanonical one exhibits states with negative specific 
heat \cite{Thirring,mkjllB,EHTa,TMHDB}, and sure enough, 
none of the computed ring pinches with negative specific heat 
minimizes the free energy functional
\begin{equation} \label{generator}
	F(\Psi)
= -{\frac{1}{8\pi}}\!
\int\! r^{-2}|\nabla \Psi|^2 d{\cal T}\!
+\! N \beta^{-1}\ln \left({\frac{2e}{N\lambda^3_{\mathrm{dB}}}}\!
\int\! e^{\beta \omega q \Psi/c}\, d{\cal T}\right) .
\end{equation}
	Actually, $F$ is unbounded below for these $\beta$, $N$ and $I$ 
values, any minimizing sequence concentrating on a singular ring
current; cf. \cite{ALYa} for a good discussion of the
translation-invariant analog.
	Of course, a real plasma would not get anywhere near such a 
singular ring current configuration, for a highly concentrated 
plasma ring is known to be susceptable to magnetofluid 
dynamical instabilities that destroy the axisymmetry.

\smallskip

\noindent{\textbf{Conclusion}}

\noindent
	To summarize, the $S$ stable ring pinches have negative specific 
heat of the gravo-thermal type and will therefore be stable {\it if and 
only if} $N$, $E$, and $I$ are essentially fixed and the toroidal invariance 
is secured, in which case the negative specific heat may aid the
ignition of thermonuclear burning.
	The ``if'' part is good news; the bad news is the ``only if'' part. 

\smallskip

\noindent
{\small{MK gratefully acknowledges financial support through a DFG 
Habilitations-Fellowship in early, and through NSF Grants \# DMS-9623220 
and \# DMS-0103808 in later stages of this project. TN gratefully 
acknowledges support by an ESA Research Fellowship in early, and 
by a PPARC Advanced Fellowship in later stages of this project. 
Sincere thanks go to S. Goldstein for so many illuminating conversations.}}

%
%
%%%%%%%%%%%%%%%%%%%%%%%%%
%   now the references. 
%%%%%%%%%%%%%%%%%%%%%%%%%%
%

\newpage
\begin{figure}
\centering\leavevmode
\epsfxsize=0.4\textwidth
\epsfbox{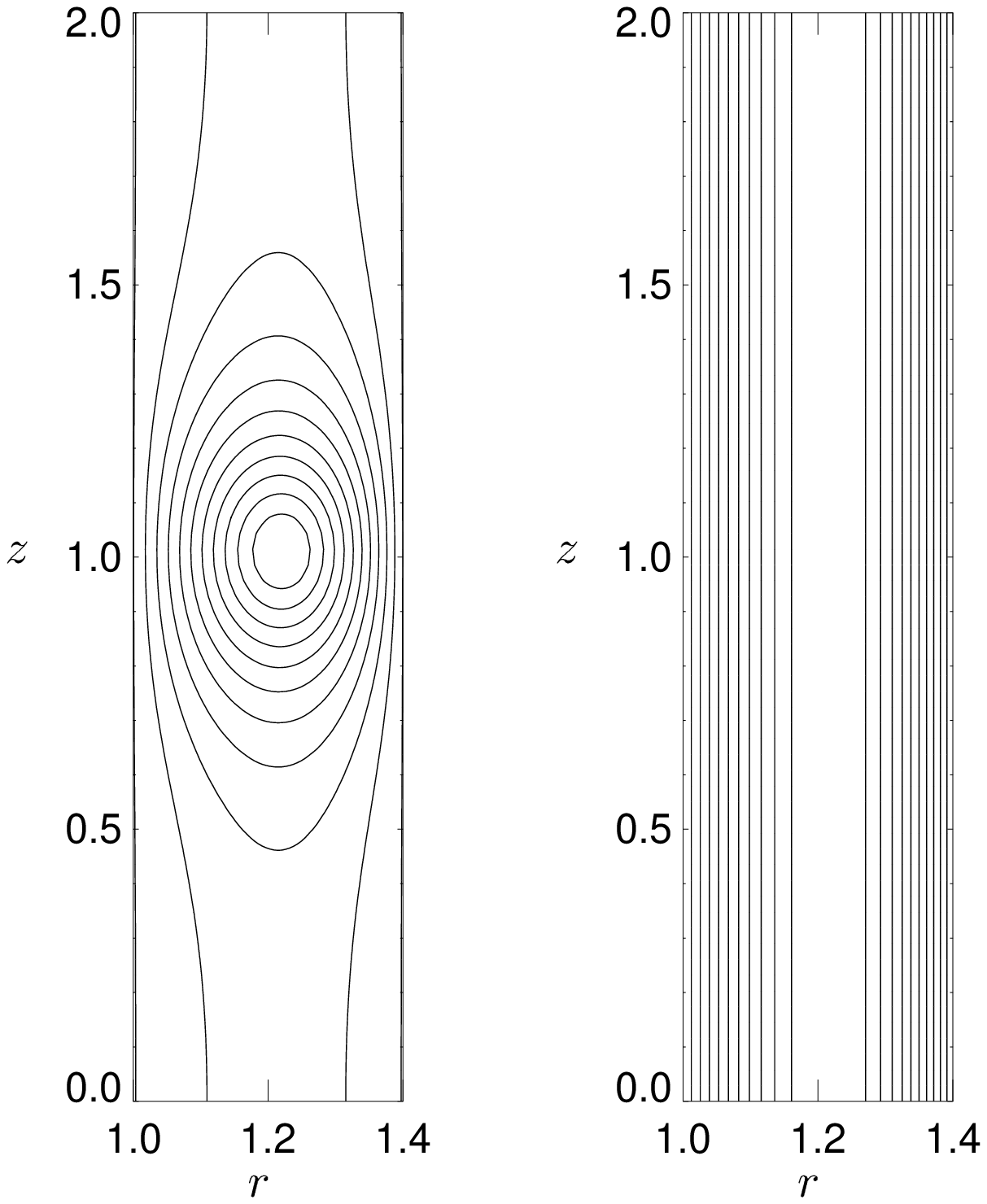}

\medskip
\caption{Poloidal magnetic lines of force of maximum entropy solutions near 
	the second order phase transition at $E_1 = 2.72 \, W_\bullet$. 
	Ring pinch (left): $W(n)= 2.34 \, W_\bullet$, 
				$\beta = 0.29 N/W_\bullet$, $z_0 = 0$;
	Sheet pinch (right): $W(n)= 3.00 \, W_\bullet$, 
					$\beta = 0.30 N/W_\bullet$.
	The toroidal hoop effect is neatly visible.\hfill}
\label{blines}
\end{figure}

\bigskip

\begin{figure}%[b]
\centering\leavevmode
\epsfxsize=0.45\textwidth
\epsfbox{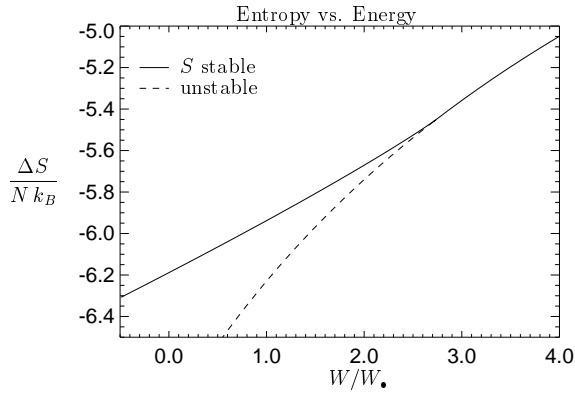}

\medskip
\caption{${\triangle}S(n)$ versus $W(n)$ for toroidal sheet pinch
	(concave branch) and  ring pinch (convex branch). }
\label{svsw}
\end{figure}

\end{document}